\documentclass[aps,nofootinbib,superscriptaddress,floatfix]{revtex4}
\usepackage{mathpazo}
\usepackage[latin9]{inputenc}
\usepackage{amsmath}
\usepackage{amssymb}
\usepackage{graphicx}
\usepackage[unicode=true, bookmarks=false, breaklinks=false,pdfborder={0 0 1},backref=section,colorlinks=false]{hyperref}
\hypersetup{colorlinks,bookmarksopen,bookmarksnumbered,linkcolor=red,urlcolor=magenta,citecolor=blue,pdfstartview=FitH}
\usepackage{color}\usepackage{textcomp}\usepackage{amsfonts}\usepackage{graphics}\usepackage{epstopdf}\usepackage{slashed}\usepackage{multirow}\usepackage{adjustbox}\usepackage{lipsum}\usepackage{rotating}
\usepackage[justification=raggedright, margin=5mm,font=sf,small,labelfont=bf, format=plain]{caption}
\definecolor{rosso}{cmyk}{0,1,1,0.4}
\definecolor{rossos}{cmyk}{0,1,1,0.55}
\definecolor{rossoc}{cmyk}{0,1,1,0.2}
\definecolor{blu}{cmyk}{1,1,0,0.3}
\definecolor{blus}{cmyk}{1,1,0,0.6}
\definecolor{bluc}{cmyk}{1,1,0,0.1}
\definecolor{verde}{cmyk}{0.92,0,0.59,0.25}
\definecolor{verdec}{cmyk}{0.92,0,0.59,0.15}
\definecolor{verdes}{cmyk}{0.9,0,0.5,0.4}

\begin{document}
\title{\color{verde} A Natural Scotogenic Model for Neutrino Mass \& Dark
Matter}
\author{Amine Ahriche}
\email{ahriche@sharjah.ac.ae}

\affiliation{Department of Applied Physics and Astronomy, University of Sharjah,
P.O. Box 27272 Sharjah, UAE.}
\affiliation{The Abdus Salam International Centre for Theoretical Physics, Strada
Costiera 11, I-34014, Trieste, Italy.}
\affiliation{Laboratoire de Physique des Particules et Physique Statistique, Ecole Normale Superieure, BP 92 Vieux Kouba, DZ-16050 Algiers, Algeria.}

\author{Adil Jueid}
\email{adil.hep@gmail.com}
\affiliation{Department of Physics, Konkuk University, Seoul 05029, Republic of Korea.}

\author{Salah Nasri}
\email{snasri@uaeu.ac.ae}
\affiliation{Department of physics, United Arab Emirates University, Al-Ain, UAE.}
\affiliation{The Abdus Salam International Centre for Theoretical Physics, Strada Costiera 11, I-34014, Trieste, Italy.}
\begin{abstract}
In this letter, we propose an extension of the scotogenic model where
singlet Majorana particle can be dark matter (DM) without the need
of a highly suppressed scalar coupling of the order $O(10^{-10})$.
For that, the SM is extended with three singlet Majorana fermions,
an inert scalar doublet, and two (a complex and a real) singlet scalars,
with a global $Z_{4}$ symmetry that is spontaneously broken into
$Z_{2}$ at a scale higher than the electroweak one by the vev of
the complex singlet scalar. In this setup, the smallness of neutrino
mass is achieved via the cancellation between three diagrams a la
scotogenic, a DM candidate that is viable for a large mass range;
and the phenomenology is richer than the minimal scotogenic model. 
\end{abstract}
\maketitle

\section{Introduction}

\label{sec:introduction}

Various astrophysical and cosmological observations indicate the existence
of a weakly or super-weakly interacting particle and which its density
constitutes about $85\%$ of matter in the universe. On the other
hand, the data from neutrino oscillations imply that neutrinos have
a tiny mass, more than six order of magnitudes lighter than than the
electron. Hence, understanding the origin of neutrino mass and the
nature of dark matter (DM) are among the strongest motivations for
going beyond the standard model (SM) of particle physics. In particular,
one hope to address these two puzzles within the same framework.

A natural explanation for the smallness neutrino mass, $m_{\nu}$,
is via the seesaw mechanism where the hierarchy between the electron
and neutrino masses is due to the hierarchy between the electroweak
(EW) scale and the singlet Majorana fermion, a new degree of freedom
that is added to the SM. However, for Yukawa couplings of order unity,
the Majorana fermion mass is many order of magnitude larger than the
EW scale, making it impossible to probe at the current and near future
high energy colliders. In addition, within this scenario there is
no DM candidate as the lightest Majorana fermion is unstable. An attractive
alternative is the radiative neutrino mass mechanism in which neutrinos
are massless at tree level and acquire a naturally small Majorana
mass term at loop level~\cite{Zee:1980ai,Zee:1985id,Babu:1988ki,Ma:1998dn,Ma:2006km,Krauss:2002px,Aoki:2008av,Gustafsson:2012vj,Nomura:2016seu,Nomura:2016ezz}
(see~\cite{Cai:2017jrq,Boucenna:2014zba} for a review). In such
models neutrino masses are calculable and their smallness follows
from Loop suppression factors and products of the Yukawa couplings.
Consequently, the new degrees of freedom that couple to the SM particles
can be of the order of the electroweak scale, and hence will potentially
be accessible at high energy colliders. Furthermore, many radiative
neutrino mass models naturally provide dark matter candidate which
itself plays a central role in generating small mass for neutrino.

The simplest realization of this neutrino-mass generation mechanism
is provided by the scotogenic model~\cite{Ma:2006km}, which is a
minimal extension of the SM by an inert scalar doublet and three singlet
Majorana fermions. In this framework, the neutrinos gets their small
masses at the one-loop level. Besides, the model allows for two possible
candidates for DM: the lightest Majorana fermion, or the lightest
neutral scalar in the inert dark doublet \footnote{The phenomenology of the scotogenic model has been extensively studied
in the literature~\cite{Ahriche:2017iar,Kitabayashi:2018bye,Borah:2018rca,Ahriche:2018ger,Mahanta:2019gfe,Hugle:2018qbw,Abada:2018zra,Baumholzer:2019twf,Mahanta:2019sfo,Das:2020hpd,Borah:2020wut}.}, which is highly constrained by the DM direct detection experiments.

However, in order to generate tiny neutrino mass in the minimal scotogenic
model, one needs either to make the Yukawa coupling in the new interactions
extremely small, or to enforce a mass-degeneracy between the $CP$-odd
and the $CP$-even scalars. Moreover, all the new Yukawa couplings
are renormalized multiplicatively~\cite{Bouchand:2012dx}, and hence
small new Yukawa couplings values are stable against radiative corrections
and will remain small along the renormalization group flow. Similar
feature can noticed for the coupling $\lambda_{5}$, contrary to the
couplings $\lambda_{3}$ and $\lambda_{4}$, if it chosen to be zero
at the electroweak scale, it remains so at higher energy scale.

In the case of scalar DM candidate, the right amount of the relic
density can be achieved only by considering the co-annihilation effect~\cite{LopezHonorez:2006gr}\footnote{Here, the scalar DM in the minimal scotogenic model is potentially
indistinguishable from inert Higgs doublet model at high energy colliders
and DM direct detection experiments.}, or/and via the assistance of the lightest Majorana fermion decay~\cite{Sarma:2020msa}.
In addition, to avoid the constraints on the DM-Nucleus scattering
cross section, we must have a suppression on the Higgs-DM
coupling $\lambda_{L}=\lambda_{3}+\lambda_{4}\pm\lambda_{5}$. However,
the coupling $\lambda_{5}$ does not need to be suppressed in this
case, and therefore the mass degeneracy between the CP-even and CP-odd
is not sharp. Consequently, the required small values of the new Yukawa
couplings make this case less interesting for both collider and LFV
experiments. This makes the scalar DM case not the most attractive
option for the scotogeneic model.

On the other hand, the lightest singlet Majorana fermion is an interesting
alternative since constraints from direct detection experiments do
not significantly affect the model parameters due to the fact that
spin-independent cross section $\sigma_{\mathrm{SI}}$ is induced
at the one-loop order~\cite{Jueid:2020yfj}. However, in this case,
the small Yukawa couplings imply that (i) the relic density is above
the Planck measurement since the annihilation cross section becomes
extremely small, (ii) the lepton flavor violating (LFV) decays occur
with extremely small branching ratios (many orders of magnitude smaller
than the experimental bounds), and hence it is almost hopeless for
the model to be probed at experiments searching for such processes,
(iii) the DM-Nucleus scattering cross section is below the neutrino
floor for most regions of the parameter space, making it impossible
to detect it in DM direct detection experiments. Therefore, for Majorana
DM, the only viable scenario is to enforce a mass-degeneracy between
the $CP$-odd and the $CP$-even scalars. Thus, it is the aim of this
paper to provide a natural explanation for the smallness of neutrino
mass without imposing highly suppressed values of $\lambda_{5}$ and,
therefore, addressing the problems of DM and LFV with non-suppressed
new Yukawa couplings which could be interesting at colliders. To do
so, we extend the scotogenic model with two singlet scalar fields;
one real and one complex; and transform under a global $Z_{4}$ symmetry
that is spontaneously broken by the complex singlet scalar vev at
an energy scale much higher than the EW scale. Due to this high scale,
the mixing of the complex singlet scalar with neutral inert can be
safely neglected, and therefore one can study the effective interactions
model after the $Z_{4}$ breaking. This is a minimal extension of
the model while retaining the concept of the scotogenic mechanism,
without suffering from the aforementioned issues for most of the space
parameters. Another extension of the minimal scotogenic model by a
real scalar field that mixes with the CP-even field without imposing
any discrete symmetry larger than $Z_{2}$~\cite{Beniwal:2020hjc}.
In this case, the new singlet modifies both the phenomenology of neutrino
masses, relaxes the constraints from the scalar DM relic density,
and opens up a large portion of the parameter space than the original
model.

The paper is organized as follow. In section~\ref{sec:model}, we
present the model, and its parameters. In section~\ref{sec:neutrino},
we derive the expression of the neutrino mass within the model. Section~\ref{sec:constraints}
is devoted to the theoretical and the experimental constraints on
the model parameter space. Then, in section~\ref{sec:DM} we study
the DM phenomenology of the model assuming the DM candidate to be
a Majorana fermion. In section~\ref{sec:collider}, we briefly discuss
the collider signatures of this model in hadronic colliders pointing
out the main differences between this model and the mininal scotogenic
one. We conclude in section~\ref{sec:conclusion}.

\section{The Model}

\label{sec:model}

We extend the SM by an inert Higgs doublet denoted by $\Phi$, three
singlet Majorana fermions $N_{i}$, a real singlet scalars $S$; and
a complex scalar $\chi$. Their quantum numbers under the $SU(3)_{c}\otimes SU(2)_{L}\otimes U(1)_{Y}$
group are depicted below 
\begin{equation}
\Phi:(1,2,1),\qquad N_{i}:(1,1,0),\qquad S:(1,1,0),\qquad\chi:(1,1,0).
\end{equation}
The Lagrangian that involves the Majorana fermions can be written
as 
\begin{eqnarray}
\mathcal{L} & \supset & h_{\alpha i}\bar{L}_{\alpha}\epsilon\Phi N_{i}+\frac{1}{2}M_{i}\bar{N}_{i}^{C}N_{i}+h.c.,\label{LL}
\end{eqnarray}
where $L_{\alpha}$ are the left-handed lepton doublets, and $\epsilon=i\sigma_{2}$
is an anti-symmetric tensor.

In this setup, the Lagrangian is invariant under a global $Z_{4}$ symmetry, that is spontaneously broken by the vacuum expectation value of the complex scalar $\left\langle \chi\right\rangle \neq 0 $ at an energy scale much higher that the electroweak scale\footnote{In this case the mixing of the CP-even part of the neutral component of the Higgs doublet with the real part of $\chi$ will be very small, and this will not affect DM relic density or the neutrino mass.}, and below this scale the Lagrangian has a residual to $Z_{2}$ symmetry. The charge assignment of the fields under both $Z_{4}$ and $Z_{2}$ are given in Table~\ref{charge}.

\begin{table}[h]
\begin{centering}
\begin{tabular}{|c|c|c|c|c|c|c|c|c|c|}
\hline 
 & $\chi$ & $S$ & $N_{i}$ & $\Phi$ & $L_{\alpha}$ & $\ell_{R\alpha}$ & $X_{SM}$\tabularnewline
\hline 
\hline 
$Z_{4}$ & $i$ & $-1$ & $-1$ & $-i$ & $i$ & $i$ & $+1$\tabularnewline
\hline 
$Z_{2}$ & $~$ & $-1$ & $-1$ & $-1$ & $+1$ & $+1$ & $+1$\tabularnewline
\hline 
\end{tabular}
\par\end{centering}
\caption{The field charges under the symmetries $Z_{4}$ at energy scale $\gg <H>$; and $Z_{2}$ around the electroweak scale (where $\chi$ decouples), where
$X_{SM}$ denotes all SM fields except the left-handed leptons $L_{\alpha}$
and the charged right-handed leptons $\ell_{R\alpha}$.}
\label{charge}
\end{table}

The most general $Z_2$-symmetric, $CP$-conserving, renormalizable, and gauge invariant potential reads\footnote{Here, the Lagrangian term $\left\{ \xi~SH^{\dagger}\Phi+h.c.\right\} $
in (\ref{eq:V}) emerges from the term $\left\{ \kappa~\chi SH^{\dagger}\Phi+h.c.\right\} $
after the $Z_{4}$ symmetry breaking, i.e., $\xi=\kappa<\chi>$.} 

\begin{eqnarray}
V\left(H,\Phi,S,\chi\right) & \supset & -\mu_{1}^{2}|H|^{2}+\mu_{2}^{2}|\Phi|^{2}+\frac{\mu_{S}^{2}}{2}S^{2}+\frac{\lambda_{1}}{6}|H|^{4}+\frac{\lambda_{2}}{6}|\Phi|^{4}+\frac{\lambda_{S}}{24}S^{4}+\lambda_{3}|H|^{2}|\Phi|^{2}\nonumber \\
 & + & \lambda_{4}|H^{\dagger}\Phi|^{2}+\frac{\omega_{1}}{2}|H|^{2}S^{2}+\frac{\omega_{2}}{2}|\Phi|^{2}S^{2}+\left\{ \xi SH^{\dagger}\Phi+h.c.\right\} ,\label{eq:V}
\end{eqnarray}
with $H$, and $\Phi$ can be parameterized as follows
\begin{equation}
H=\left(\begin{array}{c}
G^{+}\\
\frac{1}{\sqrt{2}}(\upsilon+h+iG^{0})
\end{array}\right),~\Phi=\left(\begin{array}{c}
H^{+}\\
\frac{1}{\sqrt{2}}(H^{0}+iA^{0})
\end{array}\right).\label{eq:para}
\end{equation}

The terms of the Lagrangian in (\ref{LL}) and (\ref{eq:V}) are invariant
under a global $Z_{2}$ symmetry according to the charges given in
Table~\ref{charge}. One has to mention that the assigned charges
according to $Z_{4}$ forbids the existence of the term $(H^{\dagger}\Phi)^{2}$
in the Lagrangian before or after the $Z_{4}$ breaking.

After the electroweak symmetry breaking (EWSB), we are left with three
$CP$-even scalars $(h,H_{1}^{0},H_{2}^{0})$, one $CP$-odd scalar
$A^{0}$ and a pair of charged scalars $H^{\pm}$. Their tree-level
masses are given by:

\begin{eqnarray}
m_{H^{\pm}}^{2}=\mu_{2}^{2}+\frac{\lambda_{3}}{2}\upsilon^{2},\,m_{A^{0}}^{2}=m_{H^{\pm}}^{2}+\frac{\lambda_{4}}{2}\upsilon^{2},\,m_{H_{1}^{0},H_{2}^{0}}^{2}=\frac{1}{2}\left\{ m_{S}^{2}+m_{A^{0}}^{2}\mp\sqrt{(m_{S}^{2}-m_{A^{0}}^{2})^{2}+4\xi^{2}\upsilon^{2}}\right\} \label{eq:mass}
\end{eqnarray}

with $m_{S}^{2}=\mu_{S}^{2}+\frac{\omega_{1}}{2}\upsilon^{2}$, and
$\alpha$ is the angle that diagonalises the $CP$-even mass matrix,
i.e. 
\begin{eqnarray}
\left(\begin{array}{c}
H_{1}^{0}\\
H_{2}^{0}
\end{array}\right)=\left(\begin{array}{cc}
c_{\alpha} & s_{\alpha}\\
-s_{\alpha} & c_{\alpha}
\end{array}\right)\left(\begin{array}{c}
H^{0}\\
S
\end{array}\right),\,t_{2\alpha}=\frac{4\sqrt{(m_{H_{2}^{0}}^{2}-m_{A^{0}}^{2})(m_{A^{0}}^{2}-m_{H_{1}^{0}}^{2})}}{m_{H_{1}^{0}}^{2}+m_{H_{2}^{0}}^{2}-2m_{A^{0}}^{2}},\label{eq:mixing}
\end{eqnarray}
with $c_{\alpha}\equiv\cos\,\alpha,\,s_{\alpha}\equiv\sin\,\alpha$,
and $t_{2\alpha}=\tan\,2\alpha$. Equation~\ref{eq:mass} implies
the following mass ordering 
\[
m_{H_{2}^{0}}\geq m_{A^{0}}\geq m_{H_{1}^{0}}
\]
Therefore the only possible scalar DM candidate is the light $CP$-even
scalar $H_{1}^{0}$. Besides, the decoupling limit could be achieved
only if the parameter $m_{S}^{2}$ is very large with respect to $\left|\xi\right|\upsilon$.
The model involves thirty-one additional parameters where two of them
$\mu_{1}^{2}$ and $\upsilon$ are absorbed into the definition of
the $W$-boson and the SM Higgs boson masses. The independent parameters
are chosen as follows: 
\begin{eqnarray}
\left\{ M_{i},m_{A^{0}},m_{H^{\pm}},m_{H_{1}^{0}},m_{H_{2}^{0}},h_{\alpha i},\lambda_{2},\lambda_{3},\omega_{1},\omega_{2}\right\} .
\end{eqnarray}

\section{Neutrino Mass}

\label{sec:neutrino}

\begin{figure}[h]
\begin{centering}
\includegraphics[width=0.48\textwidth]{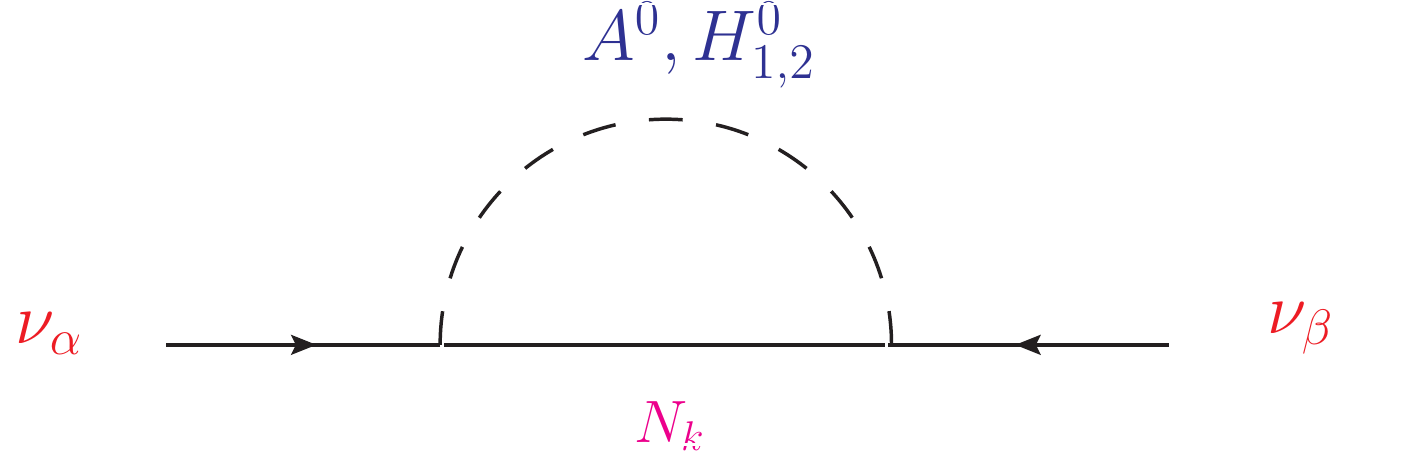} 
\par\end{centering}
\caption{Feynman diagrams responsible for neutrino mass.}
\label{fig:ms} 
\end{figure}

The neutrino mass could be generated via the three one-loop diagrams
shown in Fig.~\ref{fig:ms}.

The neutrino mass matrix elements can be written as

\begin{equation}
m_{\alpha\beta}^{(\nu)}=\sum_{k}\frac{h_{\alpha k}~h_{\beta k}~M_{k}}{16\pi^{2}}\left\{ c_{\alpha}^{2}\mathcal{F}\left(\frac{m_{H_{1}^{0}}^{2}}{M_{k}^{2}}\right)+s_{\alpha}^{2}\mathcal{F}\left(\frac{m_{H_{2}^{0}}^{2}}{M_{k}^{2}}\right)-\mathcal{F}\left(\frac{m_{A^{0}}^{2}}{M_{k}^{2}}\right)\right\} ,\label{eq:mnu}
\end{equation}

where $\mathcal{F}(x)=x\log(x)/(x-1)$. Then, according to the Casas-Ibarra
parameterization, the Yukawa couplings would have the generic form~\cite{Casas:2001sr}
\begin{equation}
h=D_{\sqrt{\varLambda^{-1}}}RD_{\sqrt{m_{\nu}}}U_{\nu}^{T},
\end{equation}
with $D_{\sqrt{\varLambda^{-1}}}=\textrm{diag}\left\{ \varLambda_{1}^{-1/2},\varLambda_{2}^{-1/2},\varLambda_{3}^{-1/2}\right\} ,\,D_{\sqrt{m_{\nu}}}=\textrm{diag}\left\{ m_{1}^{1/2},m_{2}^{1/2},m_{3}^{1/2}\right\} $,
$R$ is an arbitrary $3\times3$ orthogonal matrix, $m_{i}$ are the
neutrino mass eigenstates and $U_{\nu}$ is the Pontecorvo-Maki-Nakawaga-Sakata
(PMNS) mixing matrix; and 
\begin{align}
\Lambda_{k} & =\frac{M_{k}}{16\pi^{2}}\left\{ c_{\alpha}^{2}\mathcal{F}\left(\frac{m_{H_{1}^{0}}^{2}}{M_{k}^{2}}\right)+s_{\alpha}^{2}\mathcal{F}\left(\frac{m_{H_{2}^{0}}^{2}}{M_{k}^{2}}\right)-\mathcal{F}\left(\frac{m_{A^{0}}^{2}}{M_{k}^{2}}\right)\right\} .\label{eq:Lam}
\end{align}

In order to have an idea about the numerical values of different factors
in (\ref{eq:mnu}), one writes 
\begin{align}
\frac{m^{(\nu)}}{0.05\,\mathrm{eV}} & \backsim\left(\frac{h}{0.01}\right)^{2}\left(\frac{M_{k}}{50\,\textrm{GeV}}\right)\left(\frac{\Lambda_{k}/M_{k}}{10^{-8}}\right).
\end{align}

The smallness of the parameters $\Lambda_{k}$ in (\ref{eq:Lam}),
and therefore of the neutrino mass, can be attained within two regimes:
(1) the decoupling limit where the mixing $|\sin\,\alpha|\ll1$ is
suppressed and the heavy $CP$-even mass is too large $m_{H_{2}^{0}}\gg m_{A^{0}}$;
and (2) the quasi-degenerate masses regime $m_{H_{2}^{0}}\gtrsim m_{A^{0}}\gtrsim m_{H_{1}^{0}}$
while $|\sin\alpha|$ can take any possible value between $0$ and
$1$. Both the two regimes can be achieved with the two possible values
of the heavy $CP$-even mass that can be extracted from (\ref{eq:mass}),
which are given by 
\begin{equation}
m_{H_{2}^{0}}=m_{A^{0}}\frac{\sqrt{8+t_{2\alpha}^{2}\mp4\sqrt{4+t_{2\alpha}^{2}}}}{t_{2\alpha}}.
\end{equation}

In the decoupling limit, the neutrino mass matrix elements (\ref{eq:mnu})
can be approximated to the mininal scotogenic model formula 
\begin{equation}
im_{\alpha\beta}^{(\nu)}\simeq\frac{|\lambda_{5}|\upsilon^{2}}{16\pi^{2}}\sum_{k}\frac{h_{\alpha k}h_{\beta k}M_{k}}{\bar{m}^{2}-M_{k}^{2}}\left[1-\frac{M_{k}^{2}}{\bar{m}^{2}-M_{k}^{2}}\log\frac{\bar{m}^{2}}{M_{k}^{2}}\right]\label{eq:app}
\end{equation}
with $\bar{m}^{2}=(m_{H_{1}^{0}}^{2}+m_{A^{0}}^{2})/2$ and 
\begin{equation}
\lambda_{5}\simeq-\frac{s_{\alpha}^{2}c_{\alpha}^{2}}{\upsilon^{2}}\frac{(m_{H_{1}^{0}}^{2}+m_{H_{2}^{0}}^{2}-2m_{A^{0}}^{2})^{2}}{m_{H_{1}^{0}}^{2}+m_{H_{2}^{0}}^{2}-m_{A^{0}}^{2}}.
\end{equation}
In the quasi-degenerate masses regime ($m_{H_{2}^{0}}\gtrsim m_{A^{0}}\gtrsim m_{H_{1}^{0}}$),
the effective coupling value is given by 
\begin{equation}
\lambda_{5}\simeq-\frac{m_{A^{0}}^{2}-c_{\alpha}^{2}m_{H_{1}^{0}}^{2}-s_{\alpha}^{2}m_{H_{2}^{0}}^{2}}{\upsilon^{2}},
\end{equation}
with $\bar{m}^{2}=(c_{\alpha}^{2}m_{H_{1}^{0}}^{2}+s_{\alpha}^{2}m_{H_{2}^{0}}^{2}+m_{A^{0}}^{2})/2\simeq m_{A^{0}}^{2}$.

In the decoupling limit, by pushing the ratio $m_{H_{2}^{0}}/m_{A^{0}}$
to larger values, the ratio $\Lambda_{k}/M_{k}$ gets suppressed and
therefore the neutrino mass smallness could be easily achieved. This
scenario is much less tuned when compared with the minimal scotogenic
model, where the neutrino mass smallness is guaranteed either via
suppressed new Yukawa couplings $h_{\alpha i}$, or a suppressed value
of the coupling $\lambda_{5}=O(10^{-9})$. Moreover, the neutrino
mass smallness could be achieved without decoupling $H_{2}^{0}$ from
the mass spectrum. In this case, all the three-neutral scalars are
almost degenerate in mass $m_{H_{2}^{0}}\simeq m_{H_{1}^{0}}\simeq m_{A^{0}}$
and the mixing angle is $\alpha\simeq\pi/6$. In this case, the equal
$CP$-even contributions in (\ref{eq:Lam}) cancel the $CP$-odd one.

\section{Theoretical \& Experimental Constraints}

\label{sec:constraints}

In this section, we discuss the theoretical and the experimental constraints
which we subject the model to. These constraints are briefly discussed
below: 
\begin{itemize}
\item \textbf{Perturbativity}: all the quartic couplings of the physical
fields should be satisfy the perturbativity bounds, i.e., 
\begin{equation}
\lambda_{1,2,S},\left\vert \omega_{1,2}\right\vert ,\left\vert \lambda_{3}\right\vert ,\left\vert \lambda_{3}+\lambda_{4}\right\vert \leq4\pi.
\end{equation}
\item \textbf{Perturbative unitarity}: the perturbative unitarity must be
preserved in all processes involving scalars or gauge bosons. The
scattering amplitudes, in the high-energy limit, contains four sub-matrices
which decouple from each other due to the conservation of electric-charge,
$Z_{2}$ symmetry or $CP$-quantum numbers. We require that the eigenvalues
of these matrices to be smaller than $8\pi$. 
\item \textbf{Vacuum Stability}: the scalar potential is required to be
bounded from below in all the directions of the field space. Therefore,
the following conditions need to be fulfilled: 
\begin{equation}
\lambda_{1},\lambda_{2},\lambda_{S},\,\lambda_{1}(\lambda_{2}\lambda_{S}-\overline{\omega_{2}}^{2})-9\left(\overline{\lambda_{3}+\lambda_{4}}\right)(\lambda_{S}\left(\overline{\lambda_{3}+\lambda_{4}}\right)-3\overline{\omega_{1}}\overline{\omega_{2}})+9\overline{\omega_{1}}(3\left(\overline{\lambda_{3}+\lambda_{4}}\right)\overline{\omega_{2}}-\lambda_{2}\overline{\omega_{1}})>0,
\end{equation}
with $\overline{X}=\min\left(X,0\right)$.
\item \textbf{Gauge bosons decay widths}: The decay widths of the $W/Z$-bosons
were measured with high precision at LEP. Therefore, we require that
the decays of $W/Z$-bosons to $Z_{2}$-odd scalars is closed. This
is fulfilled if one assumes that 
\begin{equation}
m_{H_{1}^{0}}+m_{A^{0}}>M_{Z},\,m_{H^{\pm}}+m_{A_{1}^{0}}>M_{W},\,2m_{H^{\pm}}>M_{Z},\,m_{H^{\pm}}+m_{H_{1}^{0}}>M_{W}.
\end{equation}
\item \textbf{Lepton flavor violating (LFV) decays}: in this model, LFV
decay processes arise via one-loop diagrams mediated by the $H^{\pm}$
and $N_{k}$ particles as in~\cite{Toma:2013zsa,Chekkal:2017eka}.
We consider the decay branching ratios: $\mathcal{B}(\ell_{\alpha}\rightarrow\ell_{\beta}+\gamma)$
and $\mathcal{B}(\ell_{\alpha}\rightarrow\ell_{\beta}\ell_{\beta}\ell_{\beta})$
whose analytical expressions can be found in e.g.~\cite{Toma:2013zsa},
and require that are below the experimental upper bounds reported
on by \textsc{MeG} and \textsc{BaBar} experiments~\cite{Adam:2013mnn,Aubert:2009ag}. 
\item \textbf{Direct searches of charginos and neutralinos at the LEP-II
experiment}: we use the null results of neutralinos and charginos
at LEP~\cite{Abdallah:2003xe} to put lower bounds on the masses
of charged Higgs $H^{\pm}$ and the neutral scalars of the inert doublet.
The bound obtained from a re-interpretation of neutralino searches~\cite{Lundstrom:2008ai}
cannot apply to our model since the decays $A^{0}\to Z(\to\ell^{+}\ell^{-})H^{\pm}$
are kinematically forbidden. On the other hand, in most regions of
the parameter space, the charged Higgs decays exclusively into a Majorana
fermion and a charged lepton. For Yukawa couplings of order one $h_{ei}\simeq\mathcal{O}(1)$,
the following bounds are derived $m_{H^{\pm}}>100~\textrm{GeV}$~\cite{Ahriche:2018ger}.
\item \textbf{The electroweak precision tests}: in this model, the oblique
parameters acquire contributions from the existence of inert scalars.
We take $\Delta U=0$ in our analysis, the oblique parameters are
given by~\cite{Grimus:2008nb} 
\begin{eqnarray}
\varDelta T & = & \frac{1}{16\pi s_{\mathrm{w}}^{2}M_{W}^{2}}\left\{ c_{H}^{2}\,F(m_{H_{1}^{0}}^{2},m_{H^{\pm}}^{2})+s_{H}^{2}\,F(m_{H_{2}^{0}}^{2},m_{H^{\pm}}^{2})+F(m_{A^{0}}^{2},m_{H^{\pm}}^{2})\right.\nonumber \\
 & & \left.-c_{H}^{2}F(m_{H_{1}^{0}}^{2},m_{A^{0}}^{2})-s_{H}^{2}F(m_{H_{2}^{0}}^{2},m_{A^{0}}^{2})\right\} ,\nonumber \\
\varDelta S & = & \frac{1}{24\pi}\left\{ (2s_{\mathrm{w}}^{2}-1)^{2}G(m_{H^{\pm}}^{2},m_{H^{\pm}}^{2},M_{Z}^{2})+c_{H}^{2}\,G(m_{H_{1}^{0}}^{2},m_{A^{0}}^{2},M_{Z}^{2})\right.\nonumber \\
 & & \left.+s_{H}^{2}\,G(m_{H_{2}^{0}}^{2},m_{A^{0}}^{2},M_{Z}^{2})+c_{H}^{2}\,\log\left(\tfrac{m_{H_{1}^{0}}^{2}}{m_{H^{\pm}}^{2}}\right)+s_{H}^{2}\,\log\left(\tfrac{m_{H_{2}^{0}}^{2}}{m_{H^{\pm}}^{2}}\right)+\log\left(\tfrac{m_{A^{0}}^{2}}{m_{H^{\pm}}^{2}}\right)\right\} ,
\end{eqnarray}
where $s_{\mathrm{W}}\equiv\sin~\theta_{W}$, with $\theta_{W}$ is
the Weinberg mixing angle, and $F(x,y)$ and $G(x,y,z)$ are the one-loop
functions which can be found in~\cite{Grimus:2008nb}. 
\item \textbf{The signal strength $\mu_{h}^{\gamma\gamma}$}: the charged
Higgs boson can change drastically the value of the Higgs boson loop-induced
decay into two photons. The contribution of the charged Higgs boson
depends on the sign of $\lambda_{3}$; we obtain enhancement (suppression)
of $\Gamma(h\to\gamma\gamma)$ for negative (positive) values of $\lambda_{3}$~\cite{Arhrib:2015hoa}.
We use the recent measurement of $\mu_{h}^{\gamma\gamma}=\mathcal{B}(h\to\gamma\gamma)/\mathcal{B}(h\to\gamma\gamma)^{\mathrm{SM}}=1.02_{-0.12}^{+0.09}$
where we assume SM production rates for the SM Higgs boson~\cite{ATLAS:2018doi}. 
\end{itemize}

\section{Dark Matter}

\label{sec:DM}

\begin{figure}[!t]
\begin{centering}
\includegraphics[width=0.7\textwidth]{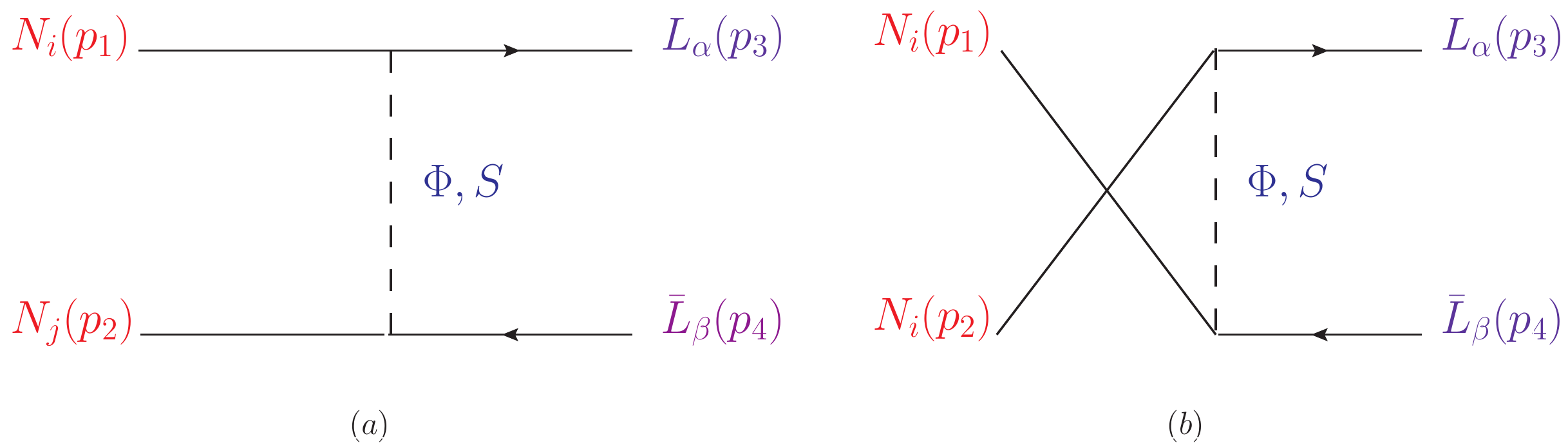} 
\par\end{centering}
\caption{DM (co-)annihilation diagrams, where $L_{\alpha}$ denotes either
charged lepton or neutrino, and $\Phi,\,S$ denote $H^{\pm}$ either
or $H_{1,2}^{0},A^{0}$.}
\label{fig:NN} 
\end{figure}

In this model, DM candidate could be either the light Majorana fermion
($N_{1}$), the light $CP$-even scalar $H_{1}^{0}$, or a mixture
of both components if they are degenerate in mass. In the case of
scalar DM, the possible annihilation channels are $W^{\pm}W^{\mp}$,~$ZZ$,~$q_{i}\bar{q}_{i}$,
$hh$, $\bar{\ell}_{\alpha}\ell_{\beta}$ and $\bar{\nu}_{\alpha}\nu_{\beta}$.
In this scenario, however, the co-annihilation effect along the channel
$H_{1}^{0}A^{0}\rightarrow X_{SM}X'_{SM}$ is important due to the
tiny mass difference $(m_{A^{0}}-m_{H_{1}^{0}})/m_{A^{0}}$. In this
case, it is favored for the couplings $h_{\alpha i}$ to be very small
due to the LFV constraints, and therefore the contribution of the
channels $\nu_{\alpha}\bar{\nu}_{\beta}$ becomes negligible, which
implies that this model with spin-$0$ DM is indistinguishable from
the usual inert doublet model. In this case of Majorana fermion as
a DM candidate, the DM self-(co-)annihilation could occurs into charged
leptons $\ell_{\alpha}^{-}\ell_{\beta}^{+}$ (light neutrinos $\nu_{\alpha}\bar{\nu}_{\beta}$)
via $t$-channel diagrams mediated by the charged scalar $H^{\pm}$
(the neutral scalars $H_{1,2}^{0},A^{0}$), as can be seen in Fig.~\ref{fig:NN}.

The relic density is given by~\cite{Srednicki:1988ce} 
\begin{equation}
\frac{\Omega_{\mathrm{DM}}h^{2}}{0.1198}\simeq\left(\frac{g_{\ast}}{100}\right)^{-1/2}\left(\frac{x_{F}}{25}\right)\left(\frac{\left\langle \sigma_{eff}\upsilon_{r}(x_{F})\right\rangle }{1.830\times10^{-9}\,\textrm{GeV}^{-2}}\right)^{-1},\label{eq:Omh}
\end{equation}
with $g_{\ast}$ is the relativistic effective degrees of freedom
that are in thermal equilibrium, $x_{F}=M_{1}/T_{F}$ is the freeze-out
parameter; and $\left\langle \sigma_{eff}\upsilon_{r}(x_{F})\right\rangle $
is the thermally averaged effective cross section at freeze-out, which
is estimated by considering the co-annihilation effect~\cite{Edsjo:1997bg}
when it's important -- see Appendix~\ref{app} for more details
--. The inverse freeze-out temperature, $x_{F}=M_{1}/T_{F}$, can
be determined iteratively by solving the transcendental equation~\cite{Srednicki:1988ce}
\begin{equation}
x_{F}=\log\left(\frac{5}{4}\sqrt{\frac{45}{8}}\frac{M_{1}M_{pl}\left\langle \sigma_{eff}\upsilon_{r}(x_{F})\right\rangle }{\pi^{3}\sqrt{g_{\ast}x_{F}}}\right).
\end{equation}

\begin{figure}[!h]
\centering \includegraphics[width=0.48\linewidth]{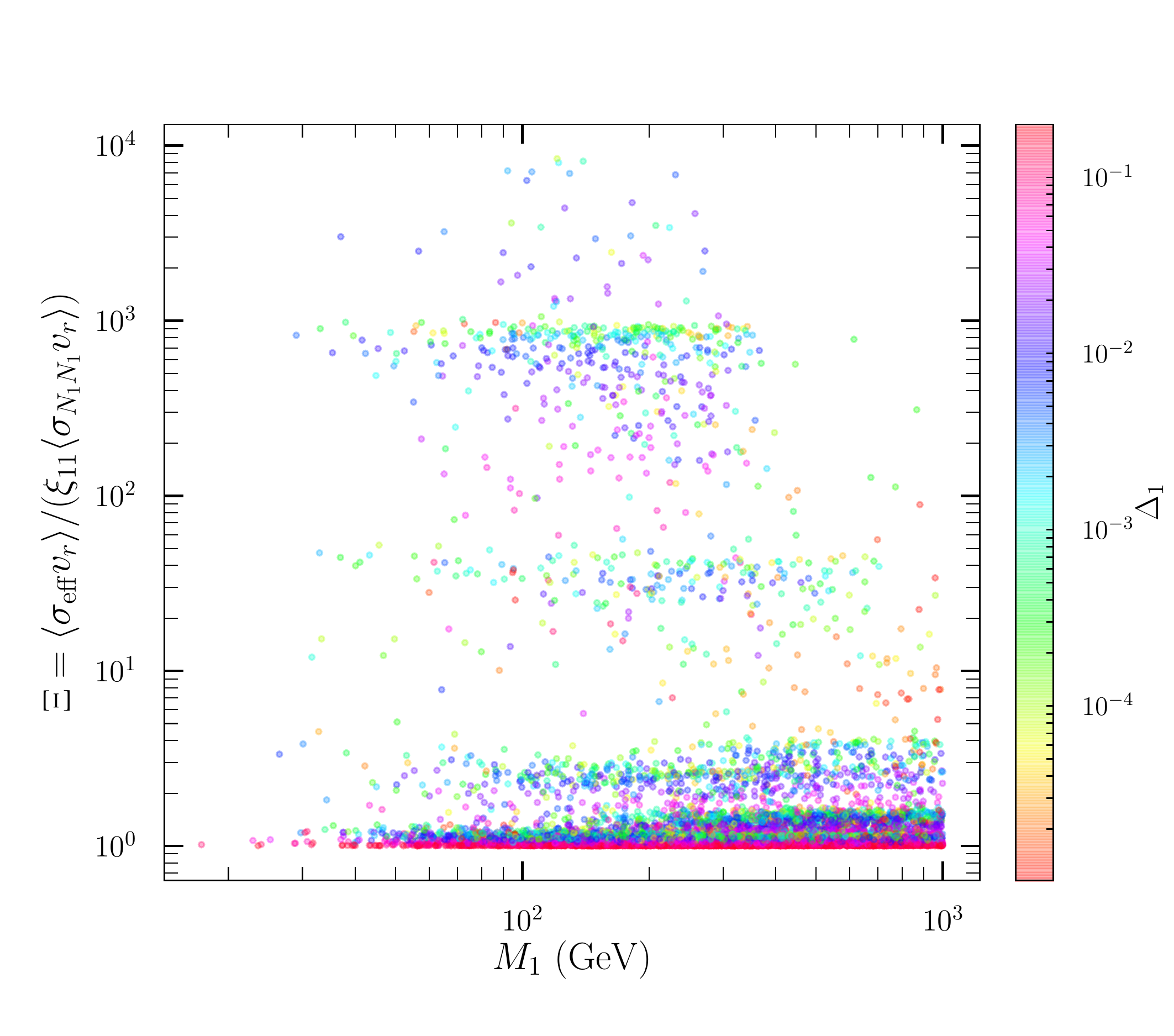} \hfill{}\includegraphics[width=0.48\linewidth]{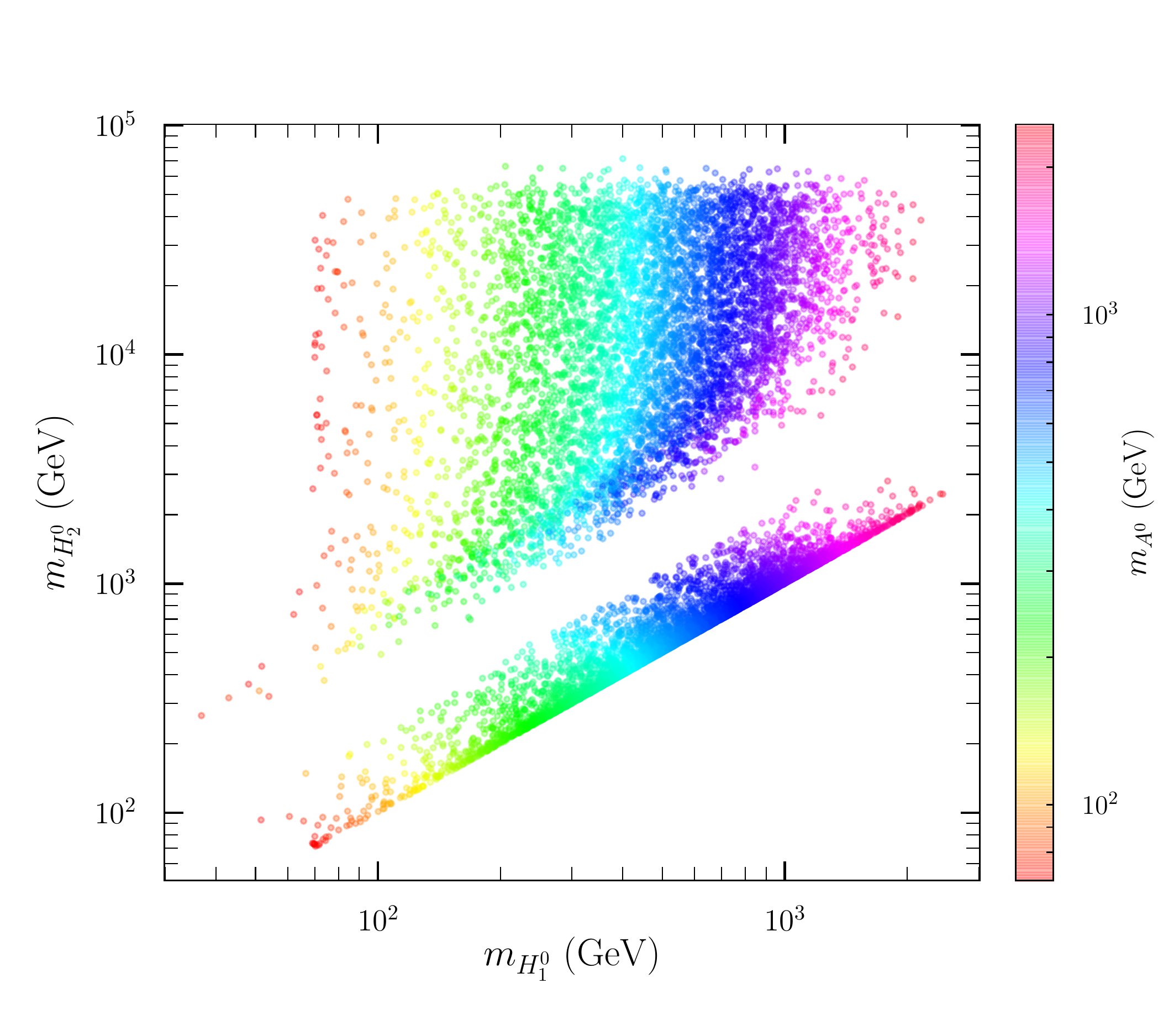}
\caption{\emph{Left:} The ratio $\Xi$ characterizing the strength of the co-annihilation
cross section as a function of the DM mass $M_{1}$ with the color
bar showing the mass splitting $\Delta_{1}=(M_{2}-M_{1})/M_{1}$.
\emph{Right:} Scatter plot on the plan of $m_{H_{1}^{0}}$ and $m_{H_{2}^{0}}$
with the palette showing $m_{A^{0}}$. The points shown in the two
panels represent randomly chosen $2\times10^{4}$ points that satisfy
constraints from relic density, direct detection, theoretical constraints,
and the electroweak precision tests.}
\label{fig:Om} 
\end{figure}

Clearly, the relic density is inversely proportional to the combination
$\Theta=\sum_{\alpha\beta}|h_{\alpha1}h_{\beta1}^{*}|^{2}$. In order
to estimate the importance of the co-annihilation effect on the annihilation
cross section, we show in Fig.~\ref{fig:Om}-right, the ratio $\Xi=\left\langle \sigma_{eff}\upsilon_{r}\right\rangle /\left(\zeta_{1,1}\left\langle \sigma(N_{1}N_{1})\upsilon_{r}\right\rangle \right)$,
where $\left\langle \sigma_{eff}\upsilon_{r}\right\rangle $ is the
full cross section within the co-annihilation effect (defined in Appendix~\ref{app});
and $\zeta_{1,1}\left\langle \sigma(N_{1}N_{1})\upsilon_{r}\right\rangle $
is its $N_{1}N_{1}$ contribution. This ratio, $\Xi$, represents
that the relative contribution to co-annihilation channels to the
annihilation one. Note that $\zeta_{1,1}=g_{eff}^{-2}$ is still depending
on the freeze-out parameter $x_{F}$. In the right panel of Fig.~\ref{fig:Om},
we display the allowed points in the model parameter space projected
on the plan of the ($H_{1}^{0}$, $H_{2}^{0}$) masses. It can clearly
be seen that there are two remarkable islands; one with quasi-degenerate
scalars and another one with $m_{H_{2}^{0}}\gg m_{H_{1}^{0}}$.

\section{Collider Signatures}

\label{sec:collider}

\begin{figure}[!t]
\centering \includegraphics[width=0.48\linewidth]{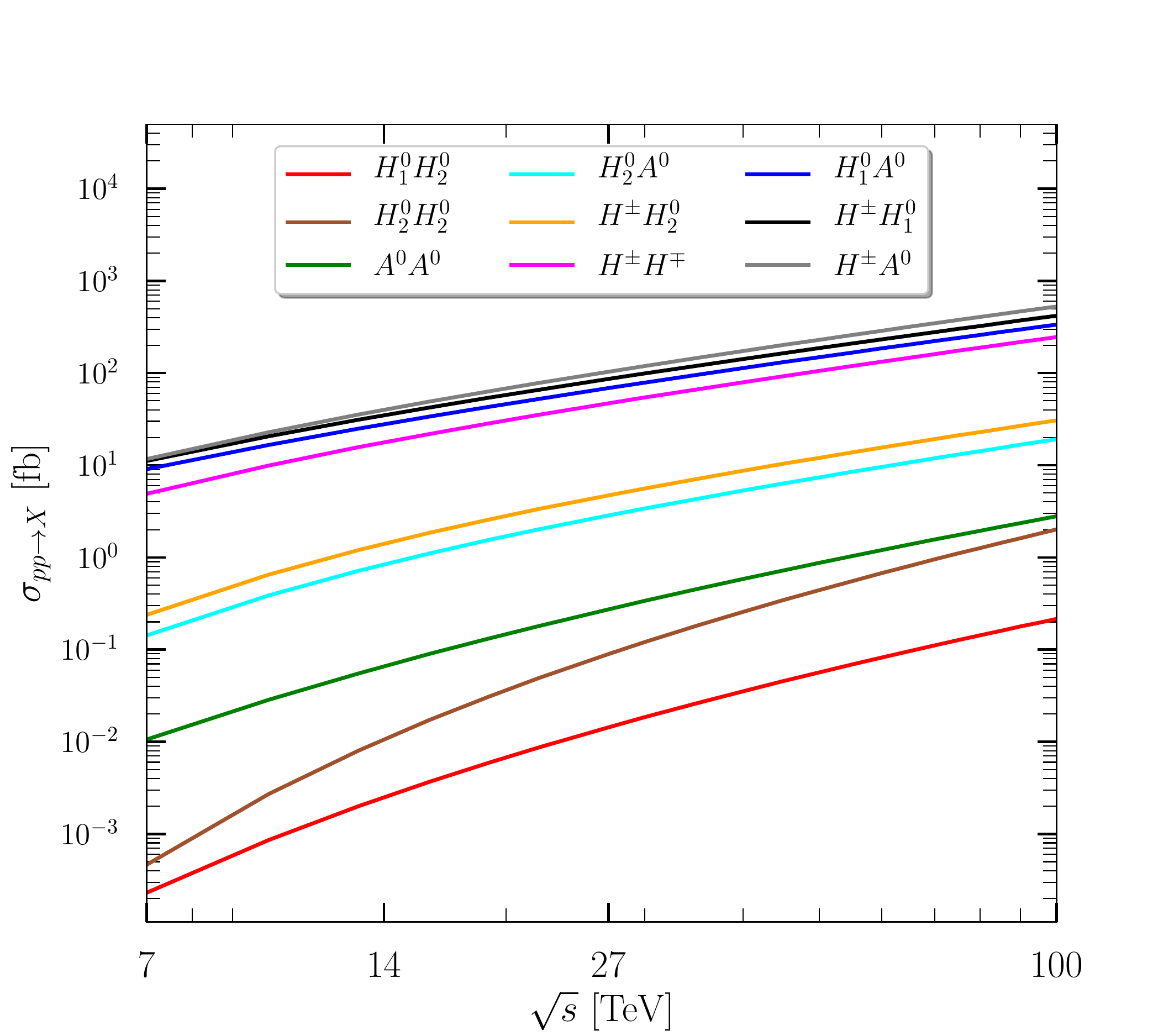}
\hfill{}\includegraphics[width=0.48\linewidth]{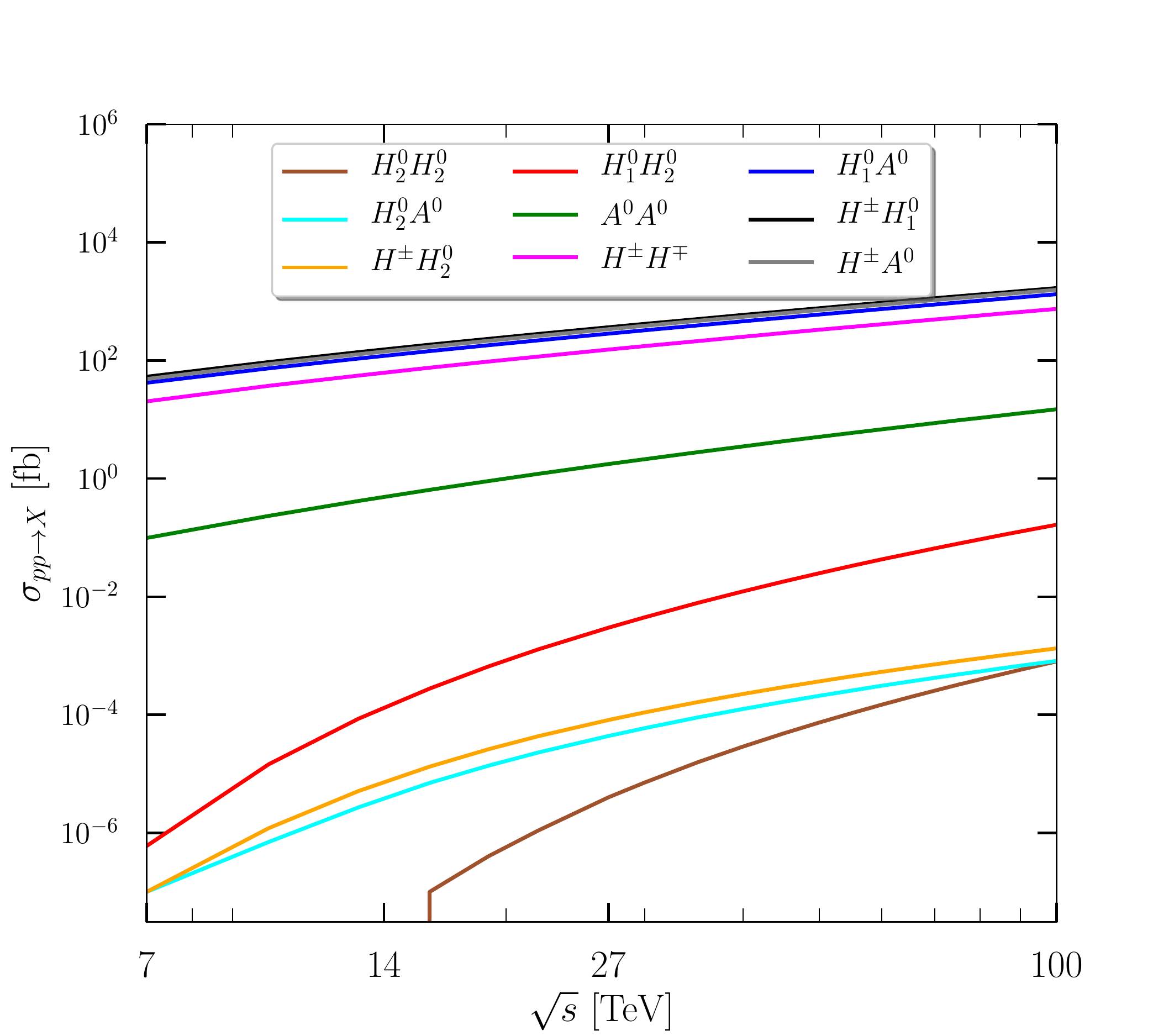}
\caption{The cross-section of pair production of inert scalars as a function
of the center-of-mass energy $\sqrt{s}$ in $pp$-collisions for \textsc{BP1}
(\emph{left}) and \textsc{BP2} (\emph{right}). The results are shown
for $H_{1}^{0}H_{2}^{0}$ (red), $H_{2}^{0}H_{2}^{0}$ (sienna), $A^{0}A^{0}$
(green), $H_{1}^{0}A^{0}$ (cyan), $H^{\pm}H_{2}^{0}$ (orange), $H^{\pm}H^{\mp}$
(rose), $H_{1}^{0}A^{0}$ (blue), $H^{\pm}H_{1}^{0}$ (black) and
$H^{\pm}A^{0}$ (gray).}
\label{fig:crosssections} 
\end{figure}

In this section, we comment briefly on the collider implications of
the model. The following discussion is based on two benchmark points
which we display in Table~\ref{tab:BPs}. In Fig.~\ref{fig:crosssections},
we display the cross sections for the pair production of the inert
scalars as a function of the center-of-mass energy $\sqrt{s}$ in
hadronic collisions\footnote{The cross sections were computed with the help of \texttt{MadGraph5\_aMC@NLO}
version 2.7.0~\cite{Alwall:2014hca} by using our \texttt{UFO} model
file~\cite{Degrande:2011ua}. We used \texttt{NNPDF30\_lo\_as\_0118}
with $\alpha_{s}(M_{Z})=0.118$~\cite{Ball:2014uwa} and renormalisation/factorisation
scales $\mu_{R,F}=1/2\sum_{i}(p_{T,i}^{2}+m_{i}^{2})$.}. In the two benchmark scenarios, the $H^{\pm}S$ production, with
$S=A^{0},H_{1}^{0}$, is dominant with a cross section of about $20$
fb ($~100$ fb) at $\sqrt{s}=14$ TeV which increases to around $400$~fb
($1$ pb) at $\sqrt{s}=100$ TeV for \textsc{BP1} (\textsc{BP2}).
In the two benchmark scenarios, $H_{1}^{0}$ and $A^{0}$ decay invisibly
into $\nu_{\ell}N_{k}$ with $100\%$ branching fraction. Moreover,
the charged Higgs boson may decay into $W^{\pm}H_{1}^{0}$ in the
first benchmark scenario although with $\mathcal{B}\simeq21.1\%$,
and the remaining decays are into $e^{\pm}N_{k}$ with branching fraction
of $70.5\%$~($100\%$) in \textsc{BP1} (\textsc{BP2}). Therefore,
the $H^{\pm}S$ production leads exclusively to a $e^{\pm}+E_{T}^{\mathrm{miss}}$
signature which can be produced in the normal scotogenic model. The
charged Higgs pair production occurs with a rate comparable to $H^{\pm}H_{1}^{0}$.
On the other hand, this process will lead exclusively to a di-electron
plus missing energy final state which also can occur in both the inert
doublet and the scotogenic models. Besides, the pair production of
neutral scalars $A^{0}H_{1}^{0}$, and $A^{0}A^{0}$ leads to an invisible
final state since both $H_{1}^{0}$ and $A^{0}$ decay invisibly.
A possibility to make these processes visible at colliders is attaching
an extra particle to the production channel such as a photon, jet
or massive gauge boson. In this case, these processes cannot be consider
smoking guns for the model as they occur in variety of scotogenic
models with comparable rates.

The processes involving the heavy $CP$-even scalar ($H_{2}^{0}$)
can be extremely important in our scenarios. There are four production
channels for $H_{2}^{0}$; $H_{1}^{0}H_{2}^{0}$, $A^{0}H_{2}^{0}$,
$H^{\pm}H_{2}^{0}$ and $H_{2}^{0}H_{2}^{0}$. The production rates
of processes involving $H_{2}^{0}$ are extremely small in the second
scenario (see right panel of Fig.~\ref{fig:crosssections}) with
cross sections of order $10^{-6}$-$10^{-4}$ fb at the HL-LHC. Consequently,
we concentrate our discussion on the first benchmark point for which
the mass of $H_{2}^{0}$ is about $349.64$~GeV. First, $H_{2}^{0}$
has four major decay channels; $hH_{1}^{0}$, $ZA^{0}$, $\nu_{\ell}N_{k}$
and $H^{\pm}W^{\mp}$ whose branching ratios are $58.7\%$, $21.56\%$,
$11.5\%$ and $7.6\%$ respectively. The $H^{\pm}H_{2}^{0}$ production
channel occurs with the highest rate going from $\simeq1$ fb at $14$
TeV to about $25$ fb at $100$ TeV. The signatures of this process
can be split according to the decays of $H_{2}^{0}$, and the subsequent
decays of the SM Higgs boson since $H_{1}^{0}$ decays invisibly with
$100\%$ branching ratio. In summary, we may have $q\bar{q}+b\bar{b}+E_{T}^{miss}$,
$2e^{\pm}+2~\mathrm{jets}+E_{T}^{\mathrm{miss}}$, $1e^{\pm}+2\ell+E_{T}^{\mathrm{miss}}$,
$1e^{\pm}+2\mathrm{jets}+E_{T}^{\mathrm{miss}}$, or $1e^{\pm}+b\bar{b}+E_{T}^{\mathrm{miss}}$.
On the other hand, $A^{0}H_{2}^{0}$ occurs with slightly smaller
production cross section than that of $H^{\pm}H_{2}^{0}$. In this
case, the $CP$-odd scalar $A^{0}$ decays invisibly and the final
states of this process consist of $h/Z^{0}(\to b\bar{b})+E_{T}^{\mathrm{miss}}$,
or $Z^{0}(\to\ell^{+}\ell^{-})+E_{T}^{\mathrm{miss}}$ among others.
Finally, the processes $H_{2}^{0}H_{2}^{0}$ and $H_{1}^{0}H_{2}^{0}$
have extremely small cross sections due to the fact they proceeds
through one-loop induced production with the exchange of the SM Higgs
boson. In summary, $H^{\pm}H_{2}^{0}$ and $A_{0}H_{2}^{0}$ are the
only processes that has rich phenomenological implications at colliders
and can be used to distinguish this model from the scotogenic model.
We close this section by noting that a more detailed analysis of the
collider implications of this model are postponed to a future study~\cite{ahriche:2020xx}.

\begin{table}[!h]
\begin{centering}
\begin{tabular}{c|c}
\hline 
\textsc{BP1} & $h_{\alpha i}=\begin{pmatrix}-0.73i & 0.068+0.021i & 0.057+0.069i\\
-8.86\times10^{-5} & -(105.27+0.64i)\times10^{-6} & (6.60-0.539i)\times10^{-6}\\
(6.60+9.72i)\times10^{-6} & 6.07\times10^{-5} & 5.07\times10^{-5}
\end{pmatrix}$ \tabularnewline
 & $m_{H_{1}^{0}}=80.72$~GeV, $m_{A^{0}}=127.89$~GeV, $m_{H_{2}^{0}}=349.64$~GeV,
$m_{H^{\pm}}=226.06$~GeV \tabularnewline
 & $M_{1}=40.93$~GeV, $M_{2}=46.37$~GeV, $M_{3}=51.94$~GeV, $\sin\alpha=0.51$ \tabularnewline
\hline 
\textsc{BP2} & $h_{\alpha i}=\begin{pmatrix}-0.85i & (8.05+2.49i)\times10^{-2} & (6.73+8.13i)\times10^{-2}\\
-5.91\times10^{-5} & (7.67-0.058i)\times10^{-4} & -(9.10+0.046i)\times10^{-4}\\
(5.05+7.44i)\times10^{-5} & 4.65\times10^{-4} & 3.88\times10^{-4}
\end{pmatrix}$\tabularnewline
 & $m_{H_{1}^{0}}=120.25$~GeV, $m_{A^{0}}=126.59$~GeV, $m_{H_{2}^{0}}=2121.75$~GeV,
$m_{H^{\pm}}=149.69$~GeV \tabularnewline
 & $M_{1}=61.49$~GeV, $M_{2}=61.86$~GeV, $M_{3}=62.49$~GeV, $\sin\alpha=0.12$ \tabularnewline
\hline 
\end{tabular}
\par\end{centering}
\caption{Benchmark points}
\label{tab:BPs} 
\end{table}

\section{Conclusion}

\label{sec:conclusion}

In this paper, we proposed a minimal extension of the so-called scotogenic
model by a real singlet that mixes with the real part of the neutral
inert component. In the minimal scotogenic model, when considering
the lightest singlet Majorana fermion to be the dark matter candidate,
the relic density requirements enforce the Yukawa couplings $h_{\alpha i}$
to be of the order $\mathcal{O}(10^{-3}\sim10^{0})$ which requires
a mass degeneracy between the $CP$-odd and $CP$-even neutral inert
components ($H^{0}$ and $A^{0}$). This degeneracy implies a suppression in the quartic couplings $\lambda_{5}$ in order to achieve the smallness
of neutrino mass. In our extension, a global symmetry $Z_{4}$-symmetry
forbids the $\lambda_{5}$ term to exist; and the mixture of the real
additional scalar with the real part of the inert component leads
to two $CP$-even eigen states with the strict mass ordering $m_{H_{1}^{0}}<m_{A^{0}}<m_{H_{2}^{0}}$.
In this setup, the neutrino mass smallness is achieved by the cancellation
between three Feynmann diagrams instead of two in the minimal scotogenic.
In consequence, we found two notable regimes on the mass spectrum
of the model: the decoupling limit $m_{H_{1}^{0}}\lesssim m_{A^{0}}\ll m_{H_{2}^{0}}$;
and the quasi-degenerate one $m_{H_{1}^{0}}\lesssim m_{A^{0}}\lesssim m_{H_{2}^{0}}$.

We further studied the impact of various theoretical and experimental
constraints on the model parameter space. We briefly discussed the
collider phenomenology of the model by displaying the cross sections
for two benchmark points corresponding to both the decoupling and
the quasi-degenerate scenarios. We found that the quasi-degenerate
scenarios are phenomenologically more interesting as they can lead
to signatures which do not appear in the mininal scotogenic model,
or in the inert doublet model. For instance, signatures with multi-jets
and multi-leptons in addition to $E_{T}^{\mathrm{miss}}$ are the
smokin-guns for these scenarios.

\acknowledgments

The authors would like to thank L. Lavoura for his valuable remarks.
The work of A. Jueid is supported by the National Research Foundation
of Korea, Grant No. NRF-2019R1A2C1009419.

\appendix
\section{The Annihilation Cross Section}\label{app}

The effective thermally averaged cross section at temperature $T=M_{1}/z$
is given by~\cite{Srednicki:1988ce} 
\begin{equation}
\left\langle \sigma_{eff}\upsilon_{r}(z)\right\rangle =\frac{z}{8M_{1}^{5}K_{2}^{2}\left(z\right)}\int_{4M_{1}^{2}}^{\infty}ds~\left\langle \sigma_{eff}(s,z)\upsilon_{r}\right\rangle \left(s-4M_{1}^{2}\right)\sqrt{s}K_{1}\left(\frac{\sqrt{s}}{M_{1}}z\right),
\end{equation}
is effective thermally averaged cross section at temperature $T=M_{1}/z$,
with $K_{1,2}$ are the modified Bessel functions.

Here, $\left\langle \sigma_{eff}\upsilon_{r}(s,z)\right\rangle $
is the effective cross section where the co-annihilation effect is
considered. Since only the Majorana fermions are considered to be
quasi-degenerate, the effective cross section at CM energy $\sqrt{s}$,
can be written as 
\begin{align}
\left\langle \sigma_{eff}(s,z)\upsilon_{r}\right\rangle & =\sum_{i,j}\zeta_{i,j}(z)\left\langle \sigma_{ij}(s)\upsilon_{r}\right\rangle ;\\
\zeta_{i,j}(z) & =\frac{(1+\Delta_{i})^{3/2}(1+\Delta_{j})^{3/2}\exp\{-(\Delta_{i}+\Delta_{j})z\}}{\left(\sum_{k}(1+\Delta_{k})^{3/2}\exp\{-\Delta_{k}z\}\right)^{2}},\label{eq:zeta}
\end{align}
with $\Delta_{k}=(M_{k}-M_{1})/M_{1}$; and $\sigma_{ij}\upsilon_{r}$
is the cross section of the processes $N_{i}N_{j}\rightarrow\ell^{+}\ell^{-},\nu\bar{\nu}$
at CM energy $\sqrt{s}$.

The relic density estimation (\ref{eq:Omh}) requires the cross section
of each annihilation channel at the centre of mass energy $\sqrt{s}$.
Here, the full amplitude of the process shown in Fig.~\ref{fig:NN}
is given by 
\begin{equation}
M=M_{(a)}-M_{(b)}=-ih_{\alpha i}h_{\beta j}^{*}\sum_{X}\left|\eta_{X}\right|^{2}\left[\frac{\bar{u}(p_{3})P_{L}u(p_{1})\bar{v}(p_{2})P_{R}v(p_{4})}{t-m_{X}^{2}}-\delta_{ij}\left(p_{1}\longleftrightarrow p_{2},t\longleftrightarrow u\right)\right],\label{eq:M}
\end{equation}
where for $\ell_{\alpha}^{-}\ell_{\beta}^{+}$ we have $\eta_{H^{\pm}}=i$,
for $\nu_{\alpha}\bar{\nu}_{\beta}$ we have $\eta_{H_{1}^{0}}=i\frac{c_{\alpha}}{_{\sqrt{2}}},\,\eta_{H_{2}^{0}}=i\frac{s_{\alpha}}{_{\sqrt{2}}},\,\eta_{A^{0}}=\frac{1}{_{\sqrt{2}}}$,
$t$ and $u$ are the Mandelstam variables; and $P_{R,L}=(1\pm\gamma^{5})/2$.
The annihilation cross section $\sigma_{ii}$ and $\sigma_{ij}\,(j\neq i)$
are given by~\cite{Lalili} 
\begin{align}
\sigma_{ii}\upsilon_{r} & =\frac{1}{32\pi s^{2}}\sum_{\alpha,\beta}\sum_{X,Y}\left|\eta_{X}\eta_{Y}\,h_{\alpha i}h_{\beta i}^{*}\right|^{2}\,\lambda(s,m_{\alpha}^{2},m_{\beta}^{2})\left\{ \mathcal{R}(Q_{X},Q_{Y},T_{+},T_{-},B)+\right.\nonumber \\
 & \left.-M_{i}^{2}(s-m_{\alpha}^{2}-m_{\beta}^{2})\mathcal{K}(Q_{X},Q_{Y},B)\right\} ,\label{eq:XS1}\\
\sigma_{ij}\upsilon_{r} & =\frac{1}{64\pi s^{2}}\sum_{\alpha,\beta}\sum_{X,Y}\left|\eta_{X}\eta_{Y}\,h_{\alpha i}h_{\beta j}^{*}\right|^{2}\,\lambda(s,m_{\alpha}^{2},m_{\beta}^{2})\mathcal{R}(Q_{X},Q_{Y},T_{+},T_{-},B),\label{eq:XS2}
\end{align}
with 
\begin{align}
\mathcal{R}(\alpha,\beta,\sigma,\theta,\eta) & =\mathcal{R}(\alpha,\beta,\sigma,\theta,-\eta)=\int_{-1}^{1}\frac{(\sigma+\eta t)(\theta+\eta t)}{(\alpha+\eta t)(\beta+\eta t)}dt,\,\mathcal{K}(\alpha,\beta,\eta)=\int_{-1}^{1}\frac{1}{(\alpha-\eta t)(\beta+\eta t)}dt.\\
\lambda(x,y,z) & =\sqrt{(x-y-z)^{2}-4yz},\,B=\frac{1}{2s}\lambda(s,m_{\alpha}^{2},m_{\beta}^{2})\lambda(s,M_{i}^{2},M_{j}^{2})\nonumber \\
Q_{X} & =\frac{1}{2}(s+2m_{X}^{2}-M_{i}^{2}-M_{j}^{2}-m_{\alpha}^{2}-m_{\beta}^{2})+\frac{1}{2s}(M_{i}^{2}-M_{j}^{2})(m_{\alpha}^{2}-m_{\beta}^{2}),\\
T_{\pm} & =\frac{1}{2}(s\pm M_{i}^{2}\mp M_{j}^{2}\pm m_{\alpha}^{2}\mp m_{\beta}^{2})+\frac{1}{2s}(M_{i}^{2}-M_{j}^{2})(m_{\alpha}^{2}-m_{\beta}^{2}).\nonumber 
\end{align}

In (\ref{eq:XS1}) and (\ref{eq:XS2}), the summations should be performed
over $\{\alpha,\beta=\ell_{\alpha}^{-}\ell_{\beta}^{+},\,X,Y=H^{\pm}\}$
and $\{\alpha,\beta=\nu_{\alpha}\bar{\nu}_{\beta},\,X,Y=A^{0},H_{1,2}^{0}\}$.

 \bibliographystyle{JHEP}
\bibliography{biblio}

\end{document}